\documentclass[12pt,draftclsnofoot,onecolumn]{IEEEtran}

\usepackage{setspace} 
\usepackage{t1enc}
\usepackage{alltt}
\usepackage{epsfig}
\usepackage{graphicx,ifthen}
\usepackage{eurosym}
\usepackage{amsfonts}
\usepackage{latexsym}
\usepackage{amssymb,amsthm,amsmath}
\usepackage{verbatim}
\usepackage{multirow}
\usepackage{bm}
\usepackage{algorithm}
\usepackage{algorithmic}
\usepackage[T1]{fontenc}
\usepackage{color}
\usepackage{ae,aecompl}

\begin{document}

\title{Towards Service-oriented 5G: Virtualizing the Networks for Everything-as-a-Service}
\author{Zheng~Chang,~\IEEEmembership{Member, IEEE,}
Zhenyu~Zhou,~\IEEEmembership{Member, IEEE,}
Sheng~Zhou,~\IEEEmembership{Member, IEEE,}
       Tapani Ristaniemi,~\IEEEmembership{Senior~Member,~IEEE,}\\
       and  Zhisheng Niu, ~\IEEEmembership{Fellow,~IEEE}
\thanks{Z. Chang and T. Ristaniemi are with University of Jyvaskyla, Department of Mathematical Information Technology, P.O.Box 35, FI-40014 Jyvaskyla, Finland. Z. Zhou is with State Key Laboratory of Alternate Electrical Power System with Renewable Energy
Sources, School of Electrical and Electronic Engineering, North
China Electric Power University, Beijing 102206, China. S. Zhou
and Z. Niu are with Department of Electronic Engineering, Tsinghua
University, 100084 Beijing, China. email: \{zheng.chang, tapani.ristaniemi\}@jyu.fi, zhenyu\_zhou@fuji.waseda.jp, \{sheng.zhou, niuzhs\}@tsinghua.edu.cn}
}
 \maketitle
\begin{abstract}
It is widely acknowledged that the forthcoming 5G architecture
will be highly heterogeneous and deployed with high degree of
density. These changes over the current 4G bring many challenges
on how to achieve an efficient operation from the network
management perspective. In this article, we introduce a
revolutionary vision of the future 5G wireless networks, in which
the network is no longer limited by hardware or even software.
Specifically, by the idea of virtualizing the wireless networks,
which has recently gained increasing attention, we introduce the
Everything-as-a-Service (XaaS) taxonomy to light the way towards
designing the service-oriented wireless networks. The concepts,
challenges along with the research opportunities for realizing
XaaS in wireless networks are overviewed and discussed.
\end{abstract}


\section{Introduction}
\label{sec:Sec1}

The future 5G will be the platform that enables the tremendous
growth of many industries, ranging from traditional wireless
networks, to the car, entertainment, manufacturing, healthcare,
and agriculture industries. It is envisioned 5G will provide a
common core to support multiple radio access technologies (RATs),
machine type communications (MTC) and many different coexisting
network and service operators. Therefore, 5G must support
convergence over traditionally separated network domains and offer
greater granularity and flexibility in control signalling and in
data transmission. Correspondingly, the architecture is expected
to be much more complex than before, in the sense that different
network entities, such as relays, small cell base stations (SBSs),
massive machines and data centers/cloud, etc, will be widely
deployed with ultra densificantion and taken as close as possible
to the end-users. Along with the rapid development of hardware
computing units, the BS in wireless networks is expected to be
deployed with powerful computing units or data centers to enable
the software defined networking (SDN) and accommodated to the
diverse service requirements. These changes, however, not only can
enable the boost of data rates, but also bring many nontrivial
challenges on how to achieve a super-efficient operation from
network management point-of-view \cite{Liang}. \par

In this light, network function virtualization (NFV) is envisioned
as one powerful tool to address these aforementioned problems in
wireless networks. In the resulted wireless network virtualization
(WNV), network infrastructures and functionalities are decoupled
from the services that they provide to maximize their utilization,
where the differentiated services can co-exist on the same
infrastructure \cite{Liang} \cite{Rost2}. What's more, due to the
fact that the network is expected to be highly heterogeneous and
extremely dense, it is natural to consider whether the network
infrastructure can be virtualized and provided to whoever wants
them and whenever they are acquired, so that the network operator
is no longer hardware-limited, nor even software-limited, in the
light of both of the hardware and software are owned by different
and dedicated network infrastructure operators. By such, every
component which used to be essential in the traditional network
management can be viewed as a service, and then can be supplied to
any (virtual) network operators/service providers (SPs) or even
directed to the end-users. Correspondingly, we refer to the
resulted system architecture as a service-oriented wireless
network with Everything-as-a-service (XaaS) which traditionally is
recognized as the service provisioning models in the cloud
computing \cite{Duan}. The new XaaS in WNV will be indeed
service-oriented, containing many new elements, such as
Data-and-Knowledge-as-a-service (DKaaS), Computing-as-a-service
(ComaaS), Radio-Access-Network-as-a-service (RANaaS),
Cache-as-a-service (CaaS) and Energy-as-a-service (EaaS), which
could be delivered over the advanced 5G infrastructure. \par

Despite the potential vision of XaaS in WNV, there are several
remaining research challenges to be addressed before its
widespread deployment, including control signalling, virtual
resource allocation, network management, and some non-technical
issues such as business model, etc. Due to the inherent random and
broadcast natures of wireless networks, these challenges need to
be tackled carefully and broadly by comprehensive research efforts
and call for a complete re-design of capabilities, architectures,
interfaces, functions, access and non-access protocols of network
services.\par

In this article, WNV is first briefly reviewed. Then by
summarizing some existed work, we discuss the XaaS taxonomy,
briefly present some definitions in XaaS and also introduce some
key enabling technologies towards the mature XaaS framework.
Challenges and research opportunities in these areas are also
discussed. This article, we hope, can attract interests from the
research and industrial communities on this emerging
interdisciplinary field, which is able to boost up the development
of the future 5G network infrastructure.

\section{Wireless Network Virtualization in 5G}

\begin{figure}[t]
\centering
\includegraphics[height=8cm, width=10cm]{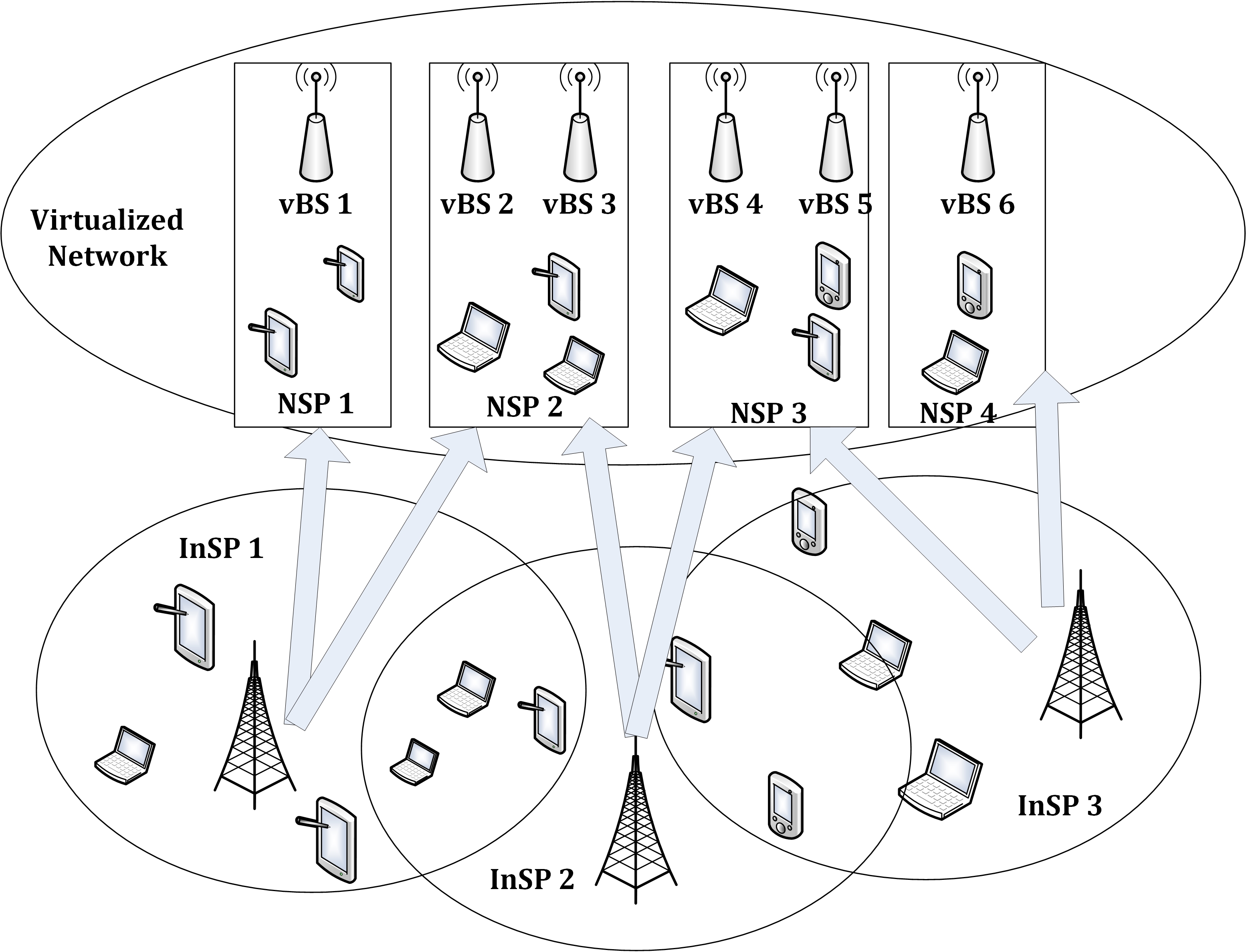}
\caption{An example of wireless network virtualization} \label{fig:example1}
\end{figure}

Virtualization have recently moved from traditional server
virtualization to wireless network virtualization. In stead of
virtualizing the computing resources in server virtualization, in
WNV, physical resources need to be abstracted to isolated virtual
resources from the infrastructure service providers (InSPs). Then,
the virtual resources can be offered to different network service
providers (NSPs). In Fig. \ref{fig:example1}, a simple
illustration of WNV is presented. In order to offer services to
the users, the NSPs in Fig. \ref{fig:example1} will ask the InSPs
about the resources. Then, the physical BSs from different InSPs
can be virtualized to virtual BSs (vBSs) and provided to different
NSPs. Hereinafter, we consider the InSPs as the ones who own the
resources, including infrastructures (hardware and software),
spectrum and many others, and refer to the NSPs as the ones who do
not have own substrate networks and need to acquire the resources
from the InSPs and provide services to end-users or other parties.
It is also worth mentioning that when the NSP acts as a reseller
or broker with respect to the resources, then naturally becomes a
InSP for the ones who buy the resources from them. Consequently, a
service-oriented wireless architecture which allows flexible and
programmable operation can be built upon the proper decoupling of
the hardware, software and radio resources. Nevertheless, the
inherit properties of the wireless communications make the problem
more complicated. Particularly, virtualizing the wireless networks
is to realize the process of radio resource virtualization,
hardware sharing, virtualization of multiple RATs \cite{Liang}.
Moreover, as the powerful computing units are becoming
indiscerptible in communications systems, virtualization of the
computing resources is an emerging option to efficiently utilize
computing units in the wireless networks \cite{Wen}.\par

\section{Everything-as-a-Service via Virtualization}

\begin{figure}[t]
\centering
\includegraphics[height=10cm, width=18cm]{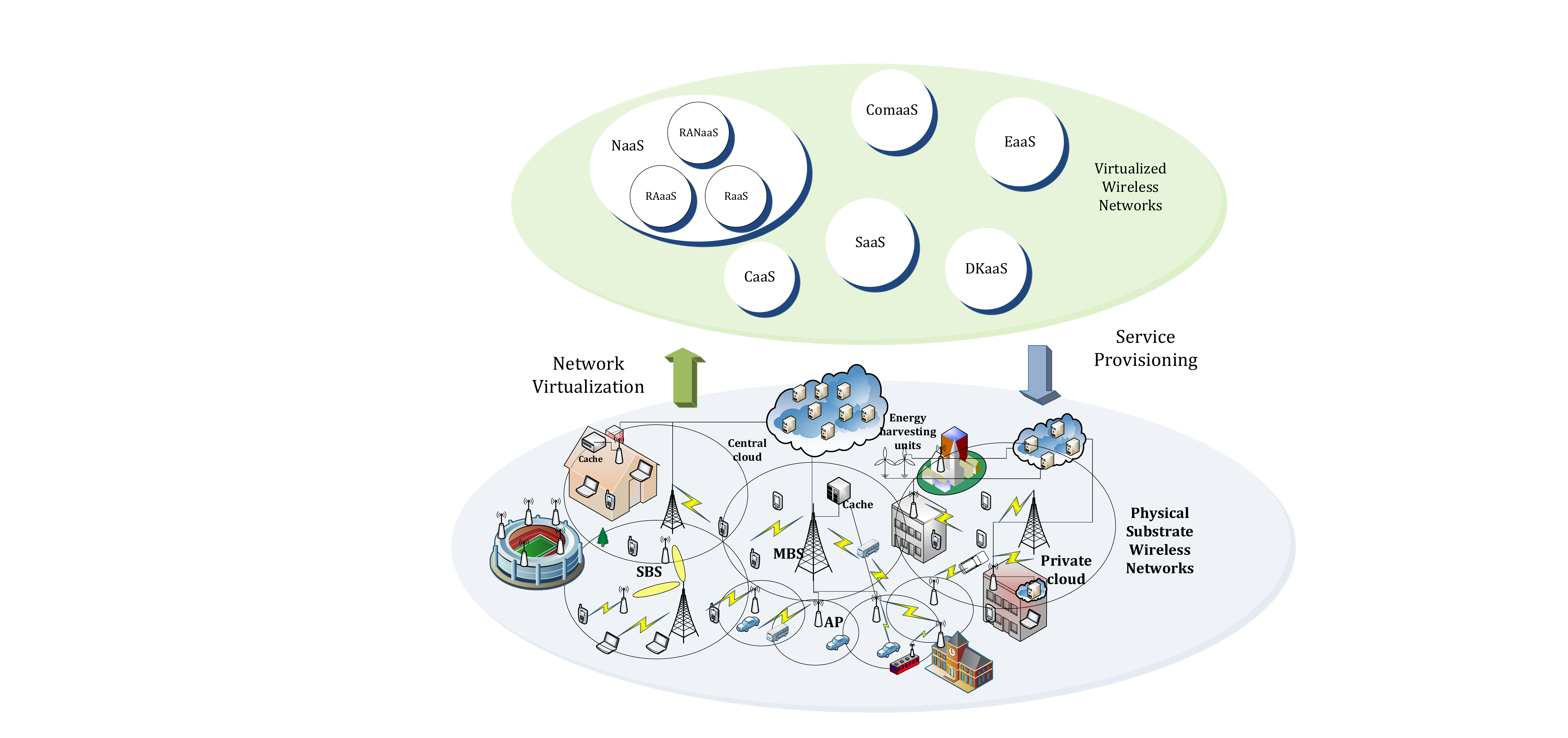}
\caption{Concept of XaaS} \label{fig:xaas}
\end{figure}

An example of XaaS is presented in Fig. \ref{fig:xaas}, where
different types of cells, such as picocell, microcell, femtocell,
and many other types of access points will be deployed and other
advances, such as cloud-RAN and energy harvesting units, will be
merged and utilized in the 5G physical substrate wireless
networks. The densitification of heterogeneous wireless networks,
together with WNV, can breakthrough the traditional obstacles on
the infrastructure and radio resources towards an efficient
network management and operation. By such, naturally, our vision
is that the network architecture will be purely service-oriented:
every component, not only BSs and spectrum, but also data,
knowledge, computing units, energy, and security, can be viewed as
a type of service that can be provided to whoever needs them and
whenever they are needed. For several years now, in the cloud
computing, researchers have been proposing and discussing many
models for defining anything "as-a-service (aaS)" \cite{Duan}.
Extracted from such a concept, we then present an XaaS taxonomy,
discuss the enabling technologies and challenges.

\subsection{Radio-Access-Network-as-a-Service (RANaaS)}

First and foremost, no matter how the network will evaluate,
it is still a radio-based network, where the wireless
infrastructure and resources are the basis for all kinds of
network operations. Therefore, the fundamental of a
service-oriented wireless system should be able to offer Radio
Access-Network-as-a-Service (RANaaS) \cite{Sabella2}.
It can be noticed that the current proposals usually
consider RANaaS is one of the products of cloud-based networks,
i.e., cloud-RAN, where all the RAN functionalities are centrally
operated. In this part, we revisit the concept of RANaaS and
further decouple the RANaaS to different categories, i.e.,
Infrastructure-as-a-Service (IaaS), Radio-Resources-as-a-Service
(RaaS) and Radio-Access-as-a-service (RAaaS), where the hardware,
software and resource can be treated separately and properly towards
a flexible and programmable 5G. \par

\subsubsection{Infrastructure-as-a-Service (IaaS)}

Unlike the current BS-dominated 4G system or behind, it is
expected that the 5G infrastructure concept will be significantly
enriched, due to the massive deployment of different types of BSs
and their ultra-density \cite{Chang1}, data caching entities,
computing centers, machines, sensors and energy harvesting units.
The explosive growth of these network elements can also
consequently increase the demands for supporting hardware, such as
backhaul, backbone and radio resource control units. In the
framework of the IaaS, each or the combination of some of these
advanced 5G features can be viewed as one kind of services offered
by the InSPs, and then can be virtualized to the NSPs or anyone
who needs them \cite{Trakas}.

\subsubsection{Radio-Access-as-a-Service (RAaaS)}
When the whole RAN is decoupled, the infrastructure and radio
access can be treated separately, which motivates the novel
concept of RAaaS, where some RAN functionalities belonging to the
protocol stack of the radio interface can be viewed as a service.
In light of RAaaS, all the software related components, such as
signalling, access and admission control, radio network
controller, gateway, and many other protocols in the RAN or core
networks, are virtualized. All of these RAN functionalities, or
part of them, may then be offered as a service by the RANaaS
platform to the NSPs, which adapts, configures, and extends their
operation to current traffic demands and keeps up with the
backhaul and access network structure requirements. \par

\subsubsection{Radio-resources-as-a-Service (RaaS)}
In the virtualized networks, after abstracting,
isolating and slicing, scalable radio resources can be better
controlled and optimized, and may be pooled independently of the
location, and transparently to the NSPs or directly to the
end-users. In the platform of RaaS, the radio resources can be
abstracted, isolated, assigned and sliced properly according to
the demands and requirements, and then are offered as a service.
By such, the available radio resources can be utilized more
efficiently by permitting different parties to share the same
spectrum.

\subsection{Data-and-Knowledge-as-a-service (DKaaS)}

The expansive wireless network is also emerging as a critical data
contributor over the air-interface, which makes us entering the
big data era \cite{Bi}.  Big data virtualization can be viewed as
one of the most valuable means through which to make sense of big
data, and thus make it more approachable to the end-users and
NSPs. Through virtualization, the big data can be abstracted,
characterized and virtualized to more valuable knowledge, which
can offer a shorter route to help decision making. \par

In this context, virtualization becomes a critical tool to convey
information in all data analysis, which also induces the the
proposal of Data-and-Knowledge-as-a-service (DKaaS). Meanwhile,
due to the fact that the delivery of large amount of data over
wireless networks truly occupies considerable amount of radio
resources, such as spectrum, power, storage, or even backhaul, the
data may prefer to be processed locally and only the necessary
information can be centrally collected, which open the
door for the third parties to join the business. Any local
organizations or even person who has the ability to collect data
or process the data can become the InSPs in this area and offer the
needed information to the NSPs. For example, the data analytic
company, spectrum broker, and smart wearable device companies can be the
DKaaS provider in this context. As such,
when DKaaS can be realized, the responsibility of carrying big
data transmission and analytic of the NSPs can be leased to
dedicated entities, and the radio resources can be better utilized
in order to obtain Quility of Service (QoS) improvement to the end-users.

\subsection{Cache-as-a-Service (CaaS)}

To improve the QoS of real-time data services and alleviate the
substantial real-time traffic on the backhaul or fronthaul,
enabling the storage and cache capabilities of BS is emerging as
one of the effective solutions \cite{Zhou}. All these features, as
we can predict, can support to realize the concept of
Cache-as-a-Service (CaaS), where cache, no matter it is either
personal or belonging to the company-own InSP, can be offered to
the NSPs. However, what prevents to realize the CaaS is its
distributed and wide deployment nature. To address such a problem,
virtualization can provide flexible and programmable virtual
caching capability to the InSPs and NSPs, in order to serve
end-users with QoS guaranteed service \cite{Li}. By such, the
content can be flexibly chunked, distributed, and stored based on
the its popularity, traffic diversity and the user demands.
\par

It is also worth noticing that besides the caching for content
delivery, cache can be applied to complement the big data
analytic. Consequently, CaaS can be merged with DKaaS, to address
the questions of matching between cache and data in the wireless
networks, i.e., problems of where, what data and when to cache
\cite{Bi}. Furthermore, due to the development of smart phone
industry, today's terminals also have large storage capacities,
which are rapidly growing but typically under-utilized. The highly
developed computing units of these devices are also capable of
processing much more complicated tasks \cite{Chang2}. To enable
the distributed cache provisioning of these devices, accurate
knowledge of the end-user demands is crucial.

\subsection{Computing-as-a-Service (ComaaS)}

In the traditional networks, the dedicated computing resources are
implemented at BS level, which resulted in a networks with an
over-provisioning of computing resources \cite{Rost2}. Such
distributed nature may prevent the full utilization of computing
resources and lead to an energy and cost inefficient networks.
Therefore, more advanced implementations should be investigated to
permit a dynamic and flexible utilization of computing resources
to network infrastructure. Utilizing similar concept as Cloud-RAN,
ComaaS emerges as am promising solution to provide immediate and
on-demand access to computing resources for the NSPs as well as
the end-users with low cost. Through the virtualization of
computing resources, ComaaS can also obtain the cost efficiency
for the InSPs, by solely utilizing the needed capacity to satisfy
NSP or user's requirement. When computing resources can be viewed
as a service, the distributed computing resources, most of which
are typically under-utilized, can be exploited as well. The
end-users are also able to contribute its computing with proper
stimulation.\par

Meanwhile, ComaaS is also one enabler for private cloud or
cloudlet, which offers hosted services to a limited number of
end-users. This is due to the fact that the use of private cloud
can be boosted by the increasing number of InSPs and NSPs. What's
more, additional private cloud expenses, including virtualization,
cloud software and cloud management tools, can also be addressed
by the ComaaS and other XaaS platforms.

\subsection{Energy-as-a-Service (EaaS)}

Energy-as-a-service (EaaS) provides a promising approach to reduce
energy costs and improve energy efficiency for both mobile users
and telecommunication operators. From the perspective of mobile
users, heavy energy consuming tasks can be offloaded to cloud
servers with unlimited computing and energy resources to fill the
gap between battery capacity limitations and high performance
expectation. Furthermore, the emerging simultaneous wireless
information and power transfer (SWIPT) technology enables mobile
terminals to "recycle" the transmit power to prolong the battery
lifetime while receiving data \cite{Chang2}. \par

On the other hand, with the advancing technologies of distributed
energy generation (DER) and distributed energy storage (DES) \cite{Zhou},
smart BSs with energy harvesting (EH) capabilities enables
operators to save excess energy in batteries and sell it back to
utility companies during peak periods, or to exploit environmental
friendly renewable energies to further reduce electricity prices
through smart energy management systems. In addition, BSs with
self-generation capabilities can compose a small-scale microgrid
and operated in islanded mode during a blackout to ensure safe and
reliable service provision. Thus, together with the emerging
energy Internet, EaaS is able to motivate the study on fundamental
relation between energy and information, and represents a novel
marketing paradigm shift from conventional passive energy
consumers to active energy prosumers.

\begin{figure}[t]
\centering
\includegraphics[height=8cm, width=10cm]{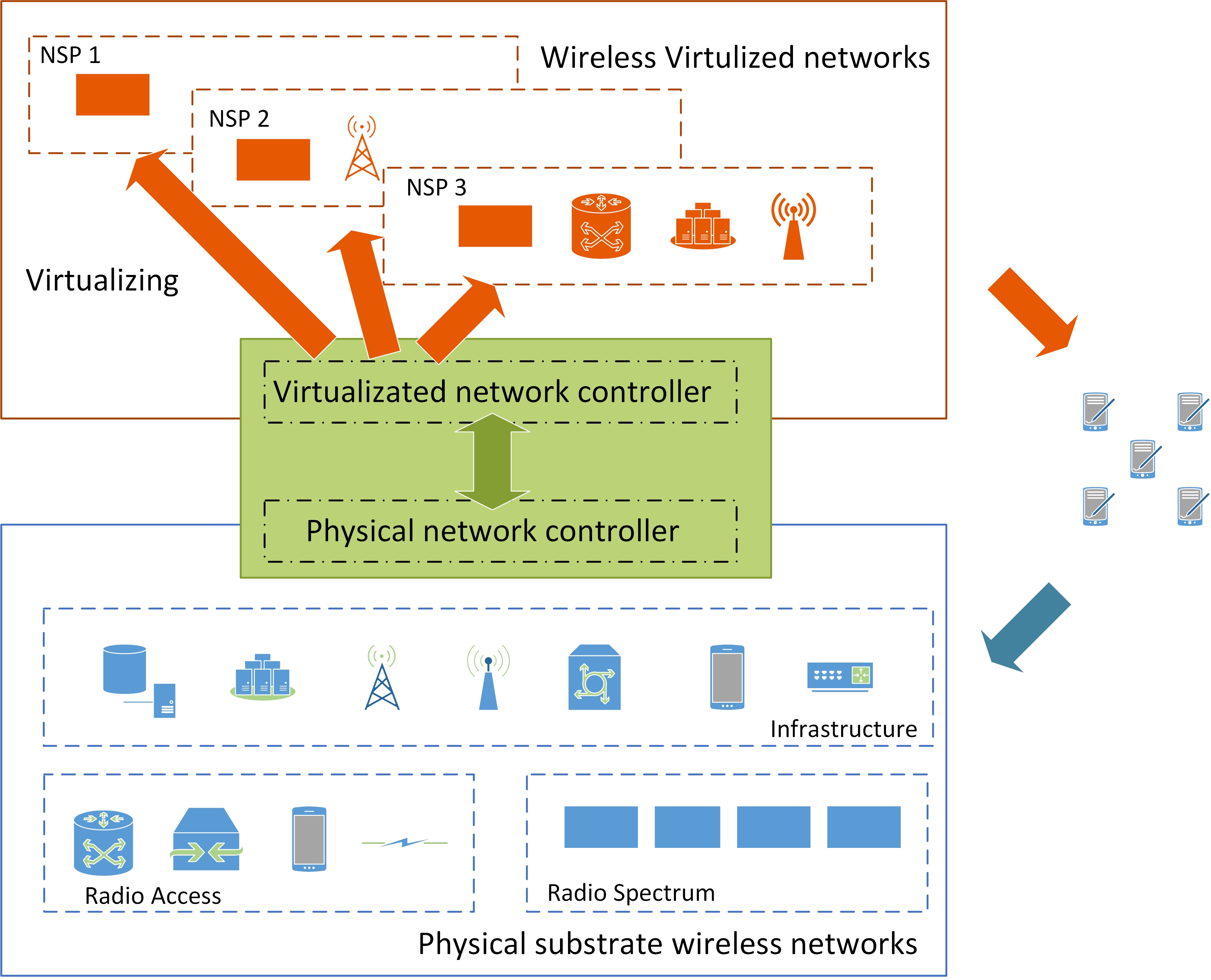}
\caption{An example of virtualizing wireless network for XaaS} \label{fig:example}
\end{figure}

\section{Challenges for enabling XaaS}
An example of virtualizing the wireless network for XaaS is shown
in Fig. \ref{fig:example}. In this example, end-users connect to
the virtual network from which they require the services, and they
also connect to the cellular network physically to obtain the
actual services. In the WNV, a physical network controller and a
virtualized network controller need to be deployed between the
virtualized wireless networks and physical wireless networks to realize the
virtualization process \cite{Liang}. We can see that enabling WNV
for XaaS confronts many challenges from the interaction of two
networks, virtual resource allocation, network infrastructure
management and involved signalling issues. In the following, we
elaborate these challenges and their impact on XaaS development.

\subsection{Signalling}
Due to the inherent broadcast nature of wireless communications
and randomness character of wireless channel, WNV is more
challenging to be realized and provided \cite{Liang}. In the
process of virtualizing the wireless network, connectivity need to
be firstly created between the NSPs and InSPs. By such, the
communications and negotiation between the NSPs and InSPs can be
established and the requirements of the NSPs for resources can be
expressed to the InSPs. In addition, in the XaaS framwork,
virtualization can happen among the InSPs as well. To facilitate
the interaction among the InSPs, a standard protocol to express
information-sharing and negotiation-handling are also necessary.
Thus, proper control signalling considering delays and reliability
needs to be explored in a careful manner to enable the
connectivity among different parties involved in wireless
virtualization. Moreover, NSPs or end-users may have different QoS
demands. Therefore, when designing the control signalling and
other overhead involved, the diversity of requirements from
different parties should be carefully treated.

\subsection{Virtual Resource Allocation}

In order to realize the XaaS in the WNV, InSPs or NSPs should
discover the available resources in the physical substrate
wireless networks. InSPs need to decide what physical resources
can be used for virtualization and NSPs can decide what resource
to choose based on the end-users' demands. Since resources will be
shared among multiple parties, an efficient resource coordination
scheme and interaction model should be investigated. Moreover, the
slicing, isolation, customization and allocation schemes are
necessary in this case, as different resources needs to be sliced
and scheduled based on the provided services to achieve a better
service differentiation against different NSPs \cite{Nikaein}.\par

In this context, the main focus of virtual resource allocation is
to realize the connection between the virtual networks and
physical networks. It includes the selection of nodes, radio
links, antenna, power, spectrum and other resources, as well as
the optimization and combination of them. Unlike wired networks,
radio resource allocation becomes much more complicated in the WNV
due to the changeability of transmission coverage, frequency
channels, user mobility, service demand, interference, transmit
power and so on \cite{Moubayed}. What's more, all the parties
involved want to maximize their own revenue, so the game theoretic
approach should be investigated to maximize the benefits in the
XaaS framework.\par

\subsection{Network Management and Deployment}
Management and deployment of the WNV are important to guarantee
the proper and efficient operation of the virtualized wireless
networks and the XaaS supported by the WNV. As the XaaS will be
based on different physical substrate networks, network management
and deployment confront many new challenges. In particularly, the
physical substrate network is usually formed by various
heterogeneous networks, each of which may have unique and specific
properties, thus, careful design for obtain the solutions for
efficient network operation and maintenance is required.\par

More specifically, in the XaaS framework, end-users should be able
to smoothly switch to the NSP from which they acquire services. In
a perfect case, the end-users should be able to access any NSP
offering the best service quality in that location. Thus, the WNV
should facilitates this mobility management through
infrastructure/resource sharing and protocols development between
the InSPs and NSPs to ensure that end-users can successfully
connect with the most appropriate NSP. Moreover, in the system
operation perspective, the WNV can require all the InSPs to share
their physical resources, which potentially allows certain InSPs
to shut down their equipment or put them into sleep when the
traffic is low. If several InSPs have overlapped coverage, or the
demand is low, it may be possible to save operation cost by
carefully choosing one of them and shut down the rest. Such system
operations are able to save the cost of NSPs as well as InSPs and
should be reconciled with virtual resource resource allocation,
isolation, and slicing, etc. The deployment of the network should
be revised as well and will be optimized based on the requirement
of the WNV and the features of the XaaS platform. For example, in
a certain area, the InSPs may need to consider how to optimally
deploy their infrastructures offer reliable services to the NSPs.
When the InSPs of certain type of XaaS are sufficient for this
area, it is not wise for them to deploy extra experiment or for
other InSPs to entering this business. Thus, the corresponding
analysis on the network management and deployment calls for
proposals from algorithmic and implementation.

\subsection{Data/Kowledge Acquisition and Abstraction}

Due to the development of data mining and processing techniques,
mobile big data is no longer viewed as a pure burden for the
wireless networks. Rather, the big data science can help the
mobile network operators to efficiently and effectively manage the
future networks with a complex architecture and provide services
to massive devices with heterogeneous demands. Meanwhile, both the
wireless and fiber-optic link have their own throughput limits,
which is considered as a inevitable bottleneck. It can be expected
that the adoption of distributed data compression and exploration
into 5G may dramatically alleviate the data transmission burden of
backhaul/fronthaul link and facilitate big data analysis in the
ultra-dense networks. Thus, how to properly acquire, process and
abstract the features of data to useful information and knowledge
are the breakthroughs on integrating the DKaaS with wireless
networks.\par

Moreover, the SP who directly serves the end-user might be the one who has
the most convinces to access the data. However, it is quite common
that the SP may not have the data processing capability nor the
data is meaningful to them. For example, the data obtained from
wireless sensor networks or wearable devices may contain extra
information that help the NSP to provide personalized and
flexible services to the end-users. Thus, how to provide these
data, compress these data or extract useful information and
knowledge from them, can attract interests from different third parties
from technique, business or social perspective are the most
challenging parts. Beside, the data/knowledge acquisition and
abstraction are also absorbing from network operation
point-of-view, as the local data processing or introducing
professionals of data mining may ease the data transmission over
wirless/wired link and abstracted knowledge can help the network
operator to run the network in a easier and cost-efficient way.
Therefore, addressing these challenges can significantly help to
realize the DKaaS concept and also open the arms of wireless
networks to embrace the upcoming big data era.

\subsection{Non-technical Challenges}
\begin{figure}[t]
\centering
\includegraphics[height=8cm, width=10cm]{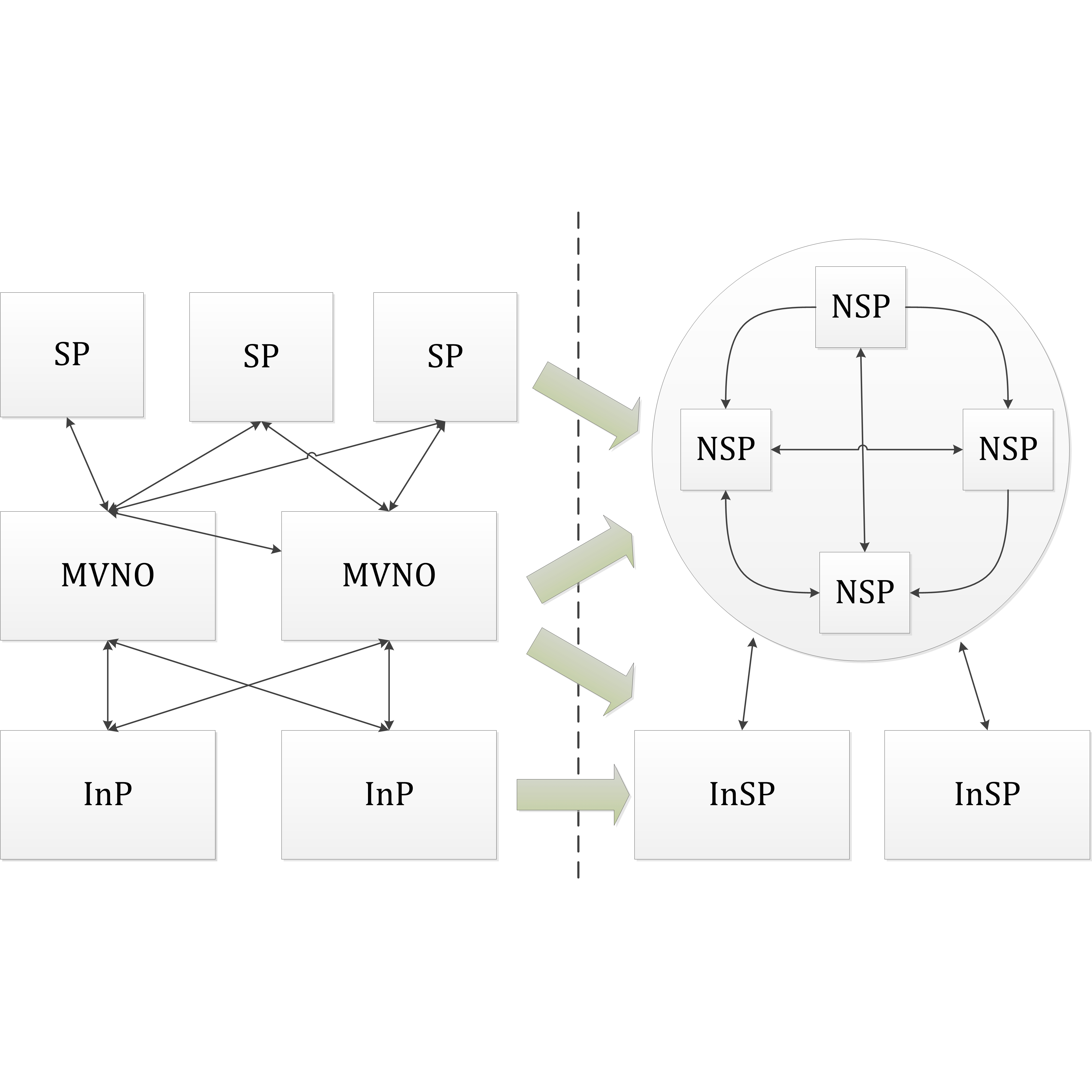}
\caption{Challenges from business model} \label{fig:business}
\end{figure}

In the technological domain, although facing aforementioned
challenges, enabling the XaaS via WNV has the great potential
gains from then network point-of-view, and then is able to provide
better services to the end-users. Besides, non-technical
challenges are brought when designing the models, such as large
volume of contextual data, massive connections, new virtual
operators, interactions between the InSPs and NSPs etc. As
presented in Fig. \ref{fig:business}, the interaction and profit
models of three layers in the WNV \cite{Liang}, i.e., service
provider (SP), mobile virtual network operator (MVNO) and INP, can
be simplified to the interaction of InSP and network service
operator (NSP). However, as the services can be decoupled and the
role of NSP can be easily changed to InSP, different involved
parties, such as service descriptions provider and service broker,
should be carefully designed and it certainly requires dedicated
and long-term research work.

\section{Future Research Direction}
While we listed some confronted research challenges, in the
following, the future research directions are presented in a
bigger picture.

\begin{figure}[t]
\centering
\includegraphics[height=8cm, width=10cm]{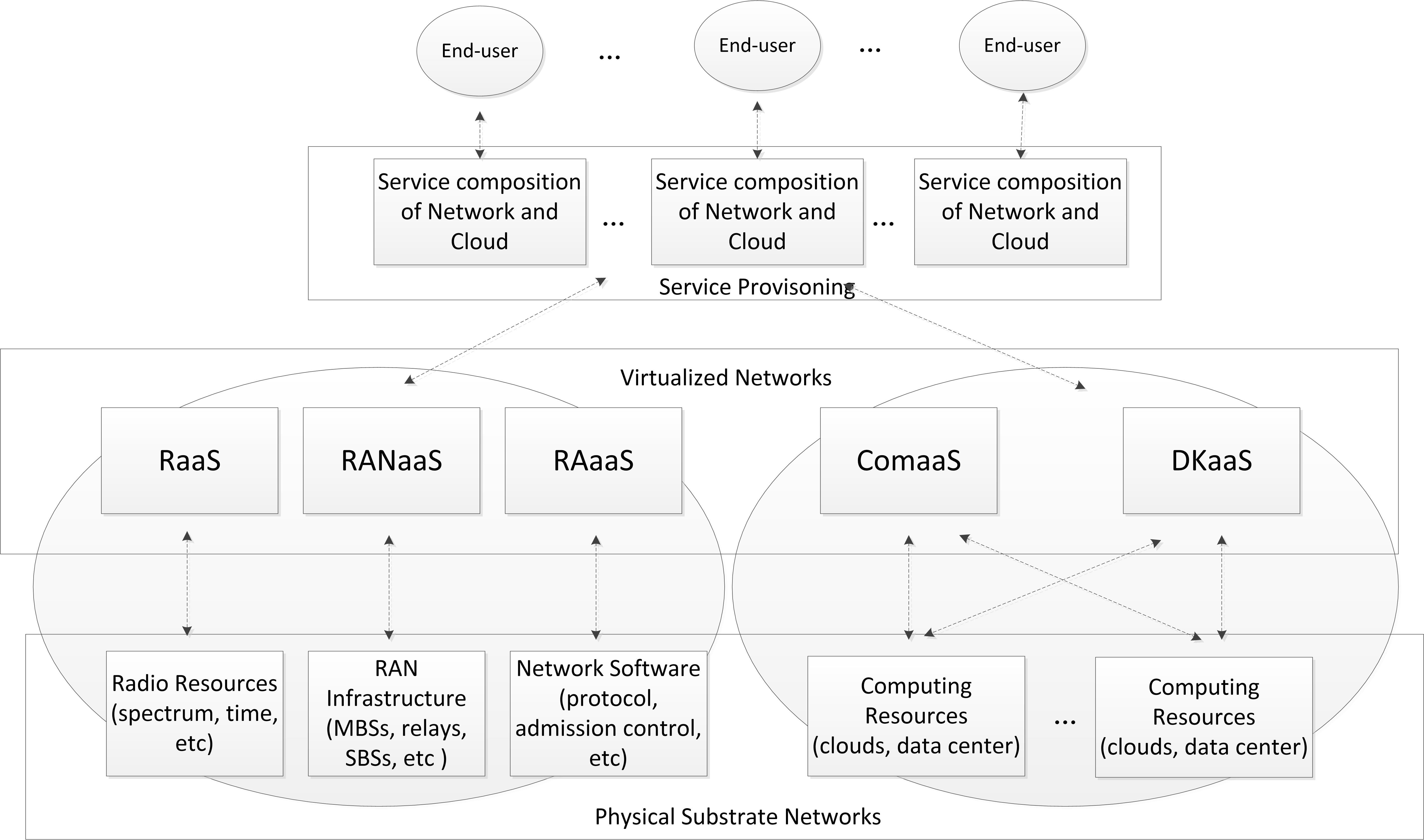}
\caption{Convergence of network and cloud}
\label{fig:cloudconvergence}
\end{figure}

\subsection{Network and Cloud Convergence}
The proposed XaaS paradigm relies on the convergence of
traditional cellular networks and cloud computing platform. From
RAN perspective, the role that cloud computing plays in networks
calls for a holistic vision that allows control, optimization and
management of both network and computing resources in a
cloud-based environment. Virtualization can be viewed as a
profound enabler for the convergence of networks and cloud. As
shown in Fig. \ref{fig:cloudconvergence}, radio resource and
computing resources can be effectively, flexibly and efficiently
virtualized into services. Then, both resulted XaaS in
communication and computing domains should be properly combined
for service provisioning to the end-users. \par

Indeed, the previous research on the RAN optimization usually
focus on the radio resource allocation, such as spectrum or power,
based on the channel state information, without considering the
computing resource and other contextual information, while most of
the research of cloud computing concentrate on the computing
resources allocation \cite{Duan1}. The limitation of previous work will
motivate the research on a joint consideration of radio and
computing resources. The challenges, in this respect, may come
from the design of a metric to measure the radio and computing
platform and to propose corresponding optimization methods. Moreover, the
protocols design between the computing resource providers and radio
resource providers in the XaaS framework also call for research
efforts. In addition, mobility issues also course challenges in a
converged network and cloud environment. Two dimensional mobility
in both physical and virtualized wireless networks, should be taken into
consideration for service provisioning. The problem can become
more serious when the mobile devices are the resources to offer
the services, which may make the service discovery and
provisioning more complex.\par

\subsection{Big Data Analytic}
Both proposed XaaS and WNV heavily rely on effective development
of big data processing technologies. At the moment, the
traditional cellular networks and recent cloud-RAN are not
designed for the incoming big data era and needs sufficient
revision to enhance the capabilities for big data analytic. To
realized the XaaS, various features of networks should be either
improved or total renovated. Exploring advanced big data
analytical tools, such as stochastic modelling to capture the
dynamic features of big data, development of data mining and
machine learning algorithms, distributed optimization and
dimension reduction. In addition, some features should be
developed and brought into the current cellular networks to future
explore their capabilities for handling vast volume of data, such
distributed caching, computing, quantization and compression,
investigation of the utilization of cloudlet and mobile cloud
processing, etc, which can help to reduce traffic amount on the
fronthaul or backhaul, also release the abundant on central data
processing units.

\subsection{Service Composition}
In order to run a network or provide QoS-guaranteed services
directly to the end-users, a NSP needs to acquire from different
InSPs in the XaaS framework. Such a inherit nature of XaaS
essentially requires to enable the service composition. For
example, when a end-user needs to watch a stream, different InSPs
may be asked from the NSPs to provide caching capability, radio
resource, infrastructure, and network functionality. Accordingly,
the proposal and investigation for energy and cost efficient
service composition will play a central role in supporting and
coordinating the XaaS framework.\par

Specifically, as loose-coupling among services is one of the
critical concepts in the XaaS and there are a large number of
services involved in the XaaS, scalability emerges as one topic
with research significance for service composition design.
Accordingly, how to take into consideration of different needs to
compose multiple services and maintain QoS requirements of
different parties are of research importance. Besides,
heterogeneity is another challenging issue to service composition.
The services that are provided to the end-users in the XaaS
platform, commonly comprise of heterogeneous services offered by
different InSPs. For example, the combination of computing and
radio resources, heterogeneous infrastructures (different types of
BSs or other network elements), each of which has its own
characteristics that may result in different technical approaches
and solutions. In addition, the on-demand, programmable and
flexible features of the XaaS framework also require dynamic and
adaptive service composition. By predicting and overseeing the
service performance and the user's satisfaction level along the
time, adaptation to QoS requirements will be beneficial for
supporting elasticity of XaaS provisioning \cite{Duan1}. Thus,
balance between system scalability, QoS awareness, user
satisfaction, and service composition is a significant research
issue.

\section{Conclusions}
As one of the main concept enables infrastructure sharing and
radio resources abstraction, wireless network virtualization
emerges as a solution to reduce operation and management expenses
of wireless networks. In this article, we introduced a
revolutionary vision of the future 5G wireless networks, in which
the network is no longer limited by hardware, radio resources or
even software. Specifically, based on the idea of virtualizing
wireless networks, the Everything-as-a-Service (XaaS) concept was
presented and elaborated to light the way towards designing a
service-oriented wireless architecture. Some important research
challenges as well as future research directions in designing the
XaaS framework were discussed and presented.


\begin{thebibliography}{1}


\bibitem{Liang}
C. Liang and F. R. Yu, "Wireless virtualization for next
generation mobile cellular networks," \emph{IEEE Wireless
Communications}, vol. 22, no. 1, pp. 61-69, Feb. 2015.

\bibitem{Rost2}
P. Rost, I. Berberana, A. Dekorsy, G. Fettweis, A. Maeder, H.Paul,
V. Suryaprakash, M. Valenti, and D. W\"ubben, "Benefits and
challenges of virtualization in 5G radio access networks,"
\emph{IEEE Communications Magazine},  vol. 53, no. 12, pp.75-82, December 2015.


\bibitem{Duan}
Y. Duan, G. Fu, N. Zhou, X. Sun, N. C. Narendra, and B. Hu,
"Everything as a service(XaaS) on the cloud: origins, current and
future trends", \emph{in proc. of IEEE 8th International
Conference on Cloud Computing}, New York, USA, June 2015.

\bibitem{Wen}
H. Wen, P. K. Tiwary, and T. Le-Ngoc, "Wireless Virtualization," Springer International Publishing,
2013.



\bibitem{Sabella2}
D. Sabella, A. De Domenico, E. Katranaras, M. A. Imran, M. Di
Girolamo, U. Salim, M. Lalam, K. Samdanis, and A. Maeder, "Energy
efficiency benefits of RAN-as-a-Service concept for a cloud-based
5G mobile network infrastructure," \emph{IEEE Access}, vol.2,
pp.1586-1597, 2014.


\bibitem{Chang1}
Z. Chang, K. Zhu, Z. Zhou, and T. Ristaniemi, "Service
provisioning with multiple service providers in 5G ultra-dense
small cell networks, " \emph{in proc. of IEEE PIMRC'15}, Hong
Kong, China, Sep. 2015.

\bibitem{Trakas}
P. Trakas, F. Adelantado, and C. Verikoukis, "A novel learning
mechanism for traffic offloading with small cell as a
service,"\emph{in proc. of 2015 IEEE International Conference on
Communications}, pp.6893-6898, London, U. K., June 2015.


\bibitem{Bi}
S. Bi, R. Zhang, Z. Ding and S. Cui, "Wireless Communications in
the era of big data," \emph{IEEE Communications Magazine}, vol.
53, no. 10, pp. 190-199, Oct. 2015.

\bibitem{Wu}
J. Wu, D. Liu, X. Huang, C. Luo, H. Cui, and F. Wu, "DaC-RAN: A
data-assisted cloud radio access network for visual
communications," \emph{IEEE Wireless Communications}, vol. 22, no.
3, pp.130-136, June 2015.


\bibitem{Zhou}
S. Zhou, J. Gong, Z. Zhou, W. Chen and Z. Niu, "GreenDelivery: proactive content caching and push with energy-harvesting-based small cells," \emph{IEEE Communications Magazine}, vol. 53, no. 4, pp. 142-149, April 2015.


\bibitem{Li}
X. Li, X. Wang, C. Zhu, W. Cai, and V. M. Leung,
"Caching-as-a-service: Virtual caching framework in the
cloud-based mobile networks," \emph{in Proc. of IEEE Infocom'15
workshop}, Hong Kong, China, April 2015.


\bibitem{Chang2}
Z. Chang, J. Gong, T. Ristaniemi and Z. Niu, "Energy efficient
resource allocation for collaborative mobile clouds with Hybrid
receivers," \emph{IEEE Transactions on Vehcular Technology}, in
press, 2016.


\bibitem{Nikaein}
N. Nikaein, E. Schiller, R. Favraud, K. Katsalis, D. Stavropoulos,
I. Alyafawi, Z. Zhao, T. Braun, and T. Korakis, "Network store:
exploring slicing in future 5G networks," \emph{in Proc. of
Proceedings of the 10th International Workshop on Mobility in the
Evolving Internet Architecture (MobiArch'15)}, Paris, France,
2015.

\bibitem{Moubayed}
A. Moubayed, A. Shami, H. Lutfiyya, "Wireless resource
virtualization with device-to-device communication underlaying LTE
network," \emph{IEEE Transactions on Broadcasting}, vol.61, no.4,
pp.734-740, Dec. 2015.

\bibitem{Duan1}
Q. Duan, Y. Yan, and A. V. Vasilakos, "A Survey on service-oriented
network virtualization toward convergence of networking and cloud
computing," \emph{IEEE Transactions on Network and Service
Management}, vol.9, no.4, pp.373-392, December 2012.




\end{thebibliography}
\end{document}